# Generation of keV hot near-solid density plasma states at high contrast laser-matter interaction.

O.N. Rosmej[1, 2], Z. Samsonova[3, 4], S. Höfer[3], D. Kartashov[3, 4], C. Arda[2], D. Khaghani[3], A. Schoenlein[2], S. Zähter[2], A. Hoffmann[3], R. Loetzsch[3,4], I. Uschmann[3,4], M.E. Povarnitsyn[5], N.E. Andreev[5,6], L.P. Pugachev[5, 6], M.C. Kaluza[3], C. Spielmann[3,4]

1. Helmholtzzentrum GSI-Darmstadt, Planckstr.1, Darmstadt, Germany
2. Goethe University, Frankfurt, Max-von-Laue-Str. 1, Frankfurt am Main, Germany
3. Institute of Optics and Quantum Electronics, Abbe Center of Photonics, University Jena, Max-Wien-Platz 1, Jena, Germany
4. Helmholtz-Institute Jena, Fröbelstieg 3, Jena, Germany
5. Joint Institute for high Temperatures of RAS, Izhorskaya st.13, Moscow, Russia
6. Moscow Institute of Physics and Technology (State University), Institutskiy Pereulok 9, Dolgoprudny, Russia

e-mail: o.rosmej@gsi.de

**ABSTRACT**

We present experimental evidence of ultra-high energy density plasma states with the keV bulk electron temperatures and near-solid electron densities generated during the interaction of high contrast, relativistically intense laser pulses with planar metallic foils. Experiments were carried out with the Ti:Sapphire laser system where a picosecond pre-pulse was strongly reduced by the conversion of the fundamental laser frequency into $2\omega$.

Complex diagnostics setup was used for evaluation of the electron energy distribution in a wide energy range. The bulk electron temperature and density have been measured using x-ray spectroscopy tools; the temperature of supra-thermal electrons traversing the target was determined from measured bremsstrahlung spectra; run-away electrons were detected using magnet spectrometers. The measured electron energy distribution was in a good agreement with results of Particle-in-Cell (PIC) simulations. Analysis of the bremsstrahlung spectra and results on measurements of the run-away electrons showed a suppression of the hot electrons production in the case of the high laser contrast.

Characteristic x-ray radiation has been used for evaluation of the bulk electron temperature and density. The measured Ti line radiation was simulated both in a steady-state and a transient approaches using the code FLYCHK that accounts for the atomic multi-level population kinetics. The best agreement between the measured and the synthetic spectrum of Ti was achieved at 1.8 keV electron temperatures and $2 \times 10^{23}$ cm$^{-3}$ electron density.

By application of Ti-foils covered with nm-thin Fe-layers we demonstrated that the thickness of the created keV hot dense plasma doesn't exceed 150nm. Results of the pilot hydro-dynamic simulations that are based on a wide-range two-temperature EOS, wide-range description of all transport and optical properties, ionization, electron and radiative heating, plasma expansion, and Maxwell equations (with a wide-range permittivity) for description of the laser absorption are in excellent agreement with experimental results. According to these simulations, the generation of keV-hot bulk electrons is caused by the collisional mechanism of the laser pulse absorption in plasmas with a near solid step-like electron density profile. The laser energy firstly deposited into the nm-thin skin-layer is then transported into the target depth by the electron heat conductivity. This scenario is opposite to the volumetric character of the energy deposition produced by supra-thermal electrons.

## I. INTRODUCTION

Laser accelerated electrons play a major role in the transfer of the laser energy into matter. Their energy distribution can usually be described by a Maxwell-like distribution function with one or more temperatures [1-4]. Depending on conditions of the laser-matter interaction, such as the laser intensity, the preplasma scale length, the polarization of the laser light, the type of the target surface etc., the fractions of "slow" electrons with energies of a few keV and "hot" electrons with relativistic energies can vary strongly [3-6]. At relativistic intensities, the interaction of the prepulse with the target on the nanosecond and picosecond time scales may lead to the target ionization and generation of the preplasma. This results into a surface modification before the arrival of the main pulse and influences its interaction with the target significantly [7, 8]. Therefore, the temporal contrast plays a decisive role in physics of the laser-matter interaction and influences the parameters of generated plasmas.

It was shown that a preplasma of a near critical electron density supports the generation of a significant fraction of MeV electrons [9-11]. Simulations of the MeV-electron spectral distribution and fraction in dependence on the preplasma scale length $L_p$ made in [12] showed dramatic increase in the hot electron temperature and conversion efficiency by increasing of the plasma scale length from $L_p=50\lambda$ up to $500\lambda$. In investigations of the electromagnetic pulses (EMP) generation made with the Ti:Sapphire laser system operating at the fundamental frequency and a low contrast [13], it was shown that relativistic run-away electrons that left the target can charge it up to 20-40 nC. Electrons, accelerated up to MeV energies in the region of a near critical electron density, easily penetrate the cold solid part of the target. As a consequence, their energy is deposited in a relatively large volume, defined by a sub-mms up to mms long stopping path. A direct hot electron current leads to a charge separation that is compensated by a return current caused by the slow bulk electrons [14]. Both collisional and resistive mechanisms lead to the target heating and creation of plasmas with temperatures that ranges from some eVs at the target rear-side and up to keV close to the target front side [1, 15, 16].

Another situation can be expected at the doubled laser frequency and high laser contrast that nowadays is available at most of laser systems. Together with the application of plasma mirrors and the frequency doubling, various techniques are used for the improvement of the amplified spontaneous emission (ASE) and the temporal contrast [17]. This allows for experiments where virtually no preplasma is generated, i.e. where the main laser pulse interacts directly with solid-state matter or with a preplasma of the scale length much shorter than the laser wavelength. This regime is of special importance for experiments on laser interaction with nano- and micro-structures [18-22] as well as for the acceleration of protons via the process of radiation pressure acceleration (RPA) that requires the application of nanometer thin foils [23-25].

In present work, we investigate parameters of plasma obtained in relativistic laser-matter interactions at high laser contrast and doubled laser frequency with planar foils. Characteristic and continuous plasma self-radiation was used for characterization of the bulk electron temperature and density as well as the temperature of the suprathermal electrons that are confined in the target by the charge separation field. The energy distribution and the charge of run-away electrons with energy above 400 keV were measured by means of a magnet spectrometer.

The paper is organized as follows: Sec. II describes the used experimental set-up; Sec. III summarizes results on measurement and simulation of the supra-thermal part of the electron energy distribution; Sec. IV presents results on measurements and evaluation of the bulk electron temperature and density of created hot plasma states using the characteristic x-ray radiation; self-consistent modeling of the laser-energy absorption and target heating is described in Sec. V followed by the summary (Sec.VI).

## II. EXPERIMENTAL SET-UP

The experiment was carried out with the Ti:Sapphire laser system JETI-40 in Jena [26]. The picosecond pre-pulse has been reduced down to $10^{-8}$ using the second harmonic generation (SHG) in a 700 μm-thick KDP crystal. The residual fundamental energy was suppressed by a factor of $10^6$-$10^7$ using a set of three dichroic mirrors. The p-polarized laser pulse at the 400nm wavelength and the energy of 180-200 mJ was focused onto the target under 45 degrees into a 5 μm spot by means of $f/3$ off-axis Al parabolic mirror. The resulting peak intensity reached $1.7\times10^{19}$ Wcm$^{-2}$ and a corresponding normalized vector potential $a_0 \approx 1.4$ The duration of the SHG pulse is estimated to be 45 fs (FWHM) and at the level of $10^{18}$ Wcm$^{-2}$ is 80 fs. Ti-foils of 25 μm thickness have been used as targets. This foil thickness was chosen in order to prevent the refluxing of electrons [2] with energies up to 100 keV.

The diagnostic setup used in this experiment comprised an x-ray spectrometer with a cylindrically bent Highly Ordered Pyrolytic Graphite (HOPG)-crystal ($2d = 6.71$Å) in the von Hamos geometry that was facing the target front side along the target normal (see Fig.1). The spectrometer was sensitive for a wide spectral window of 4.4 - 7.9 keV and provided a spectral resolution of $\lambda/\delta\lambda\sim1000$. Every spectrum was registered using BASF MS image plates (IP) and was accumulated over 10-25 laser shots. A hard x-ray detector (HXRD) based on a filter-attenuation method and a Timex detector [27] operating in the single photon counting regime were used for the analysis of the bremsstrahlung radiation with photon energies up to 0.2 MeV. HXRD was equipped with two Ti and seven Ta-filters with thicknesses from 25 up to 2000 µm and allowed for the registration of photons from 5 up to 500 keV.

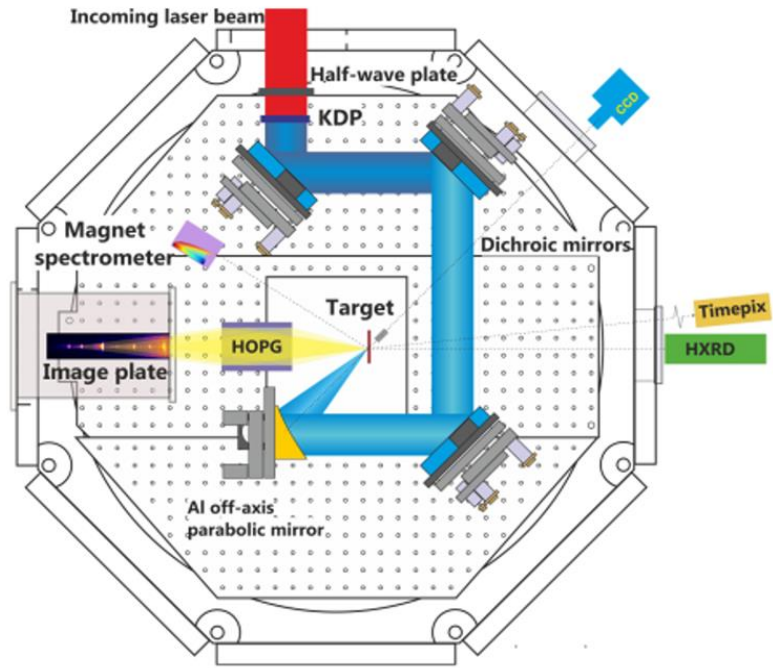

FIG. 1. Experimental set-up shows the laser beam path via the KDP-crystal, three dichroic mirrors and the off-axis parabola onto the target and the arrangement of the photons and particles diagnostics.

A Timepix detector was placed behind a 230 µm thick kapton window at a distance of 3.7 m from the target rear side and was blocked with a 200 µm-thick Al-filter in order to guarantee a single photon resolution. For the reconstruction of the measured spectra, the photon energy dependent quantum efficiency, filter and window transmission as well as absorption in the air have been taken into account.

Target bulk temperatures covering a wide parameter range from tens up to thousands of eV and the bulk electron density were determined from the analysis of the $K_\alpha$ profile broadening [1] and characteristic transitions in *Li*- and *He*-like charge states of Ti-ions. The spectral distribution of the continuum radiation was used to deduce the temperature of suprathermal electrons trapped in the target by the charge separation field. The energy spectra of electrons that escaped the target either from the target front or the rear side were measured by means of a 0.25 T magnetic spectrometer equipped with BASF MS image plates which were absolutely calibrated for electron energies from 100 keV up to 10 MeV [28].

### III. MESUREMENTS AND SIMULATIONS OF THE HOT ELECTRON DISTRIBUTION FUNCTION.

The spectral distribution of the bremsstrahlung radiation emitted by suprathermal electrons traversing the target and measured by the Timepix detector is presented in Fig.2. It can be approximated by the Maxwell-like electron energy distribution with two temperatures: electrons with $T_1 = 1.5\pm0.2$ keV provide a major input in the total radiation and electrons with $T_2 = 19.6\pm0.1$ keV contribute to 3%. Analysis of the spectra measured using the filter attenuation method results into 15-17 keV hot electron temperature what is in a good agreement with Timepix measurements. Contribution of electrons with energies above 100 keV could not be measured by applied tools since their signal was on the level of the experimental noise (see Fig.2). However, the Timepix detector is designed for detection of photons with energies above 20-30 keV that is why an additional diagnostics was used for quantitative estimate of the bulk electron temperature.

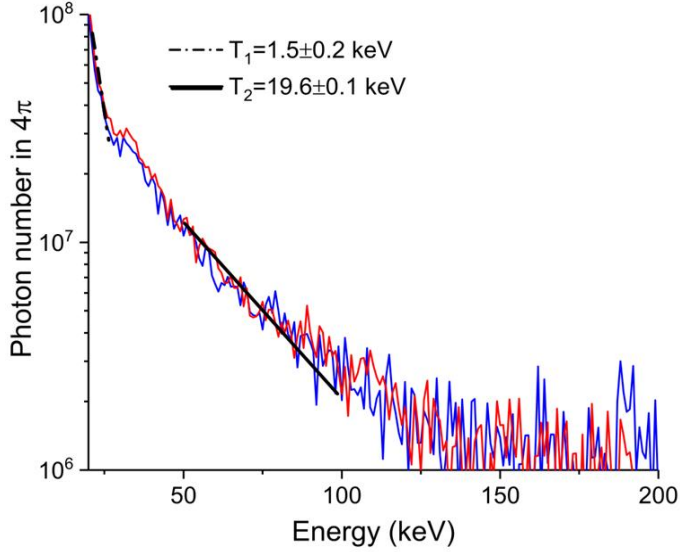

FIG. 2. Bremsstrahlung spectrum measured with the Timepix detector and fitted by the electron energy distribution with two temperatures $T_1 = 1.5\pm0.2$ keV and $T_2 = 19.6\pm10.1$ KeV.

The spectral distribution of the run-away electrons measured with of the 0.25 T magnetic spectrometer resulted in 0.41 MeV hot electron temperature what is in a good agreement with a pondermotive temperature scaling [29]. The electron spectra measured at different angles to the target normal at the front and rear target sides were rather similar. The number of run-away electrons has been estimated using measured PSL values and the calibration curves made for the BASF MS image plate [28]. Accounting for a nearly isotropic character of the laser accelerated run-away electrons with a temperature of 0.4 MeV we obtain $10^8$ particles with energies $E_e > 0.4$ MeV in $4\pi$ per laser pulse. This number of particles corresponds to the total charge of 10 pC.

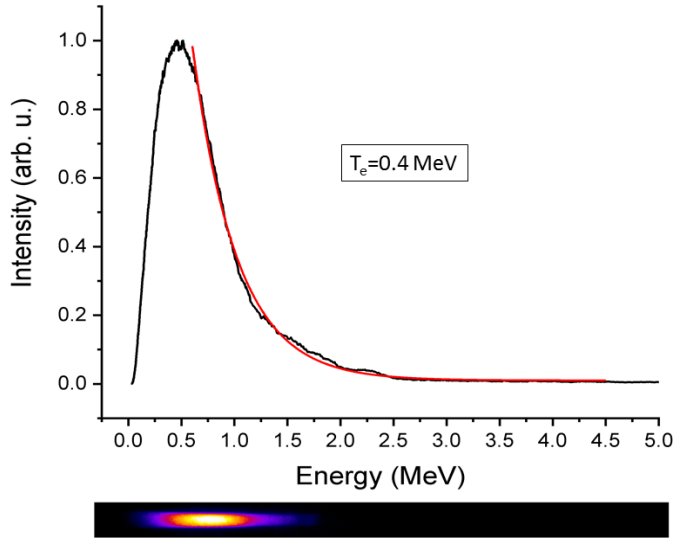

FIG. 3. Spectral distribution of the run-away electrons with estimated temperature of 0.4 MeV and $10^7$-$10^8$ particles. The spectral distribution measured at the front and rear target sides at different angles to the target normal showed rather isotropic character.

In the short pulse laser-matter interaction, where the laser energy is strongly confined in time and space, measurements are mostly time and space integrated. In order to get insight into evolution of the involved processes we performed three-dimentional (3D) Particle-in-Cell (PIC) simulations describing the electron distribution function during and after the laser pulse action. The PIC simulations were carried out with the Virtual Laser Plasma Laboratory code [40] and were performed for laser-matter interaction conditions similar to those in the experiment. Figure 4 presents energy spectra of electrons at different moments of the laser-target interaction. At the maximum of the laser intensity ($t=0$ fs) one can find the Maxwell-like electron energy distribution described by three temperatures. The fraction of electrons with $T_1 = 2-4$ keV dominates the spectra at all interaction times. Hot tails with $T_2 = 20-40$ keV and $T_3 = 420$ keV exist at the peak of the laser pulse and disappear one after another at $t=+40$ fs and $t=+80$ fs. The inset in Fig. 4 shows a projection of the momentum ($p_x$, $p_y$) at the maximum of the laser pulse intensity on target (0 fs) that has a rather isotropic character which is also preserved at later times (+40 fs, +80 fs). This simulation result is in a good agreement with nearly isotropic distribution of the run-away electrons detected in the experiment.

All three temperatures retrieved from the PIC-simulations are consistent with experimental results provided by the x-ray spectroscopy ($T_1$), the Timepix detector ($T_1$, $T_2$), the filter attenuation method ($T_2$), and the electron magnet spectrometer ($T_3$). Estimated number of electrons (~ $10^8$) with energies above 0.4- 0.5 MeV and the temperature $T_3$ obtained using results of the magnet spectrometer agrees agreement with the PIC-simulations.

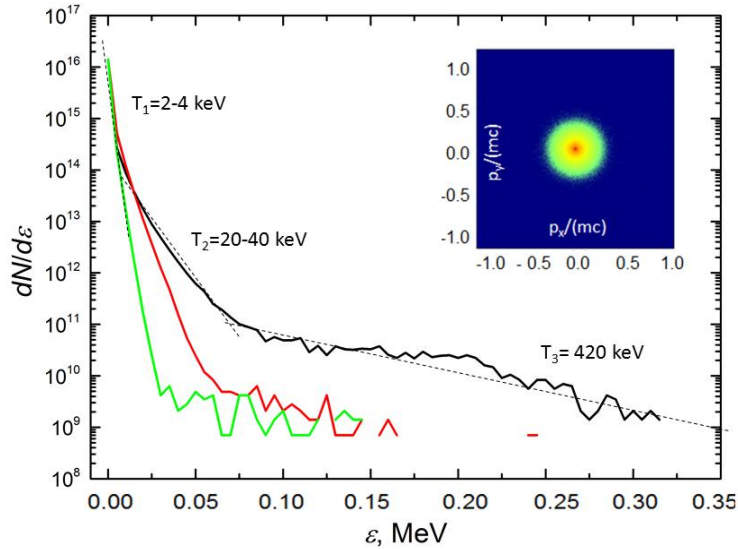

FIG. 4. Simulated electron energy spectra at different time steps during the interaction: $t=0$ fs (black curve), $t=+40$ fs (red curve), and $t=+80$ fs (green curve). The inset in Fig 4 shows the projection of the of the electron momentum space to the plane ($p_x$, $p_y$) at the peak of the pulse on target (0 fs). Its rather isotropic character is also preserved at later times (+40 fs, +80 fs).

## IV. X-RAY DIAGNOSTICS OF THE Ti-PLASMA BULK ELECTRON TEMPERATURE AND DENSITY.

For diagnostics of the bulk electron temperature and density we used characteristic $K$-shell radiation of Ti-plasmas. The $K$-shell radiation of highly charged ions is very suitable for such analysis because of the simplicity of characteristic transitions in ions with one and two bound electrons [30]. In order to ensure the appearance of highly ionized states in spectra in the case of stationary plasmas (plasma parameters are assumed to be constant in time and space), the relation between the $K$-shell ionization potential of the diagnostic element with the nuclear charge $Z_n$ and the expected electron temperature $T_e$ must be as following: $Z_n^2 \times Ry \sim 5 \times T_e$ (eV) for a Collisional Radiative Equilibrium (CRE) and $Z_n^2 \times Ry \sim 10 \times T_e$ (eV) for a Local Thermal Equilibrium (LTE); here $Ry =13.6$ eV and $Z_n$ is a nuclear charge. In the case of transient plasmas, where the electron temperature/density vary faster than the characteristic times required for collisional ionization/excitation of bound electrons, the requirement on the plasma temperature demanded to reach highly ionized atomic states is strongly coupled to the $n_e \times \tau_p$ parameter, where $n_e$ is a plasma electron density and $\tau_p$ is the plasma life-time in a hot and dense state. At values above $n_e \times \tau_p \cong 10^{11}$ - $10^{12}$ s·cm$^{-3}$ transient plasma states approach the steady state regime. For $n_e \times \tau_p < 10^{10}$ s·cm$^{-3}$, the charge state distribution of the ionized atoms will be defined not only by the electron temperature and density but will depend as well on time. In the phase when the temperature rises, ions will be less ionized, showing up lower charge states than it would be expected for given plasma parameters, while in the recombination phase, when the temperature drops, the situation is opposite.

Analyzing plasma generated at high laser intensities it has to be ensured that highly charged states are generated due to collisional ionization by free plasma electrons and not due to the optical laser field ionization. Estimations based on ADK (Ammosov, Delone, Krainov) theory [31] show that the laser intensity of $10^{19}$ Wcm$^{-2}$ is sufficient to field-ionize titanium up to the *B*-like state ($Ti^{17+}$). Further ionization to higher charge states instead will be governed by collisions with free plasma electrons. The dominant role of collisional processes that in turn depend on plasma temperature and density secures the applicability of the x-ray spectroscopy methods for the evaluation of plasma parameters.

Characteristic radiation emitted from the Ti-plasma and measured with the HOPG crystal spectrometer in the photon energy range 4.5-5 keV is shown in Fig. 5. The spectrum contains emission lines from the following radiative transitions into the *K*-shell vacancy: $K_\alpha$ and $K_\beta$– transitions from the levels with principal quantum numbers *n*=2 and *n*=3 into the *K*-shell hole of weakly ionized Ti-atoms; *K*-shell transitions of intermediate charge states with vacancies in the *M* and *L*-shell (*F*- up to *Be*-like ions) and *K*-shell transitions from one and double excited states in *Li*- and *He*-like Ti ions. While at given intensities intermediate charge states can be produced by the laser optical field and/or collisional ionization at 100-500 eV plasma temperature, excited *He*-like states can be generated only collisionaly in the dense hot plasma region with the electron temperature above ~ 1 keV. The presence of $Ti^{+20}$ in the measured spectrum gives by itself a direct evidence for high electron temperature and density that are required to reach these charge states in a sub-picosecond time scale.

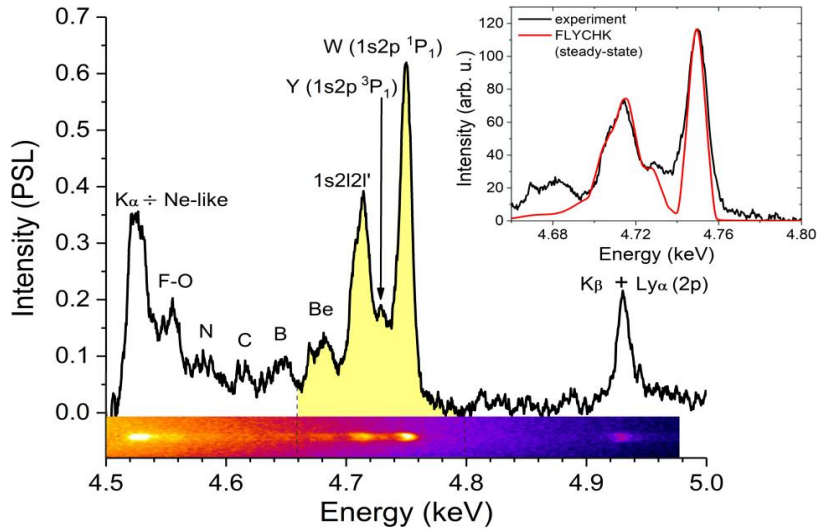

FIG. 5. *K*-shell radiation of the laser heated Ti-plasma: x-axis is the photon energy in keV, y-axis is the signal intensity on the image plate in PSL (Photo Stimulated Fluorescence). One observes the $K_\alpha$ and $K_\beta$ transitions originating in the deep target layers heated by suprathermal electrons and K-shell transitions in highly ionized Ti-ions. The inset shows the best fit of the spectral region that includes radiation of the *Be*-, *Li*- and *He*-like ions of Ti made in a steady-state approximation for $T_e$=1250 eV and $n_e$=1.7×10$^{23}$ cm$^{-3}$, line intensities are in arbitrary units.

Electrons with temperature $T_2$ = 19.6 keV, as measured by the time-pix detector, have a µm long stopping path in the target and a full divergence angle close to 180°. Consequently, their energy is spread over a relatively large volume leading to plasma temperatures from a few eVs up to few tens of eVs. *K*-shell ionization of Ti governed by suptrathermal electrons with energies far above a threshold $Z_n^2 \times Ry$ = 5-6 keV followed by a prompt radiative decay into the *K*-shell hole, gives rise to $K_\alpha$ and $K_\beta$ radiation of weakly ionized Ti-atoms (up to $Ti^{12+}$). As shown in [1], the analysis of the $K_\alpha$-profile broadening that incorporates *K*-shell transitions of weakly ionized Ti-ions up to the *Ne*-like state (partially ionized *M*-shell but occupied *L*-shell), allows to determine a plasma temperature in "warm" foil regions heated by suprathermal electrons and the compensating return current. In our case, this method gives temperatures of 20-50 eV.

Generation of the intermediate charge states (*F*- up to *Be*-like ions) can proceed as well collisionally. This requires together with a high density of free electrons the electron temperatures in the range of 100-500eV. Such temperatures exist due to the distribution of the laser intensity across the focal spot and into the target depth. The "visualization" of these charge states with partially occupied L-shell in the x-ray spectrum is possible only if at the same time the *K*-shell vacancies will be produced due to the impact ionization by hot electrons that propagate across the laser spot and deep into the target [32, 33].

The relative intensities of the *He*-like resonance transition $W$ ($1s^2\ ^1S_0 - 1s2p\ ^1P_1$) and dielectronic satellites ($1s^2 2l - 1s2l2l'$, $l, l' = s, p$) originating from the double excited states of *Li*-like ions are sensitive to the bulk electron temperature [34, 35]. The higher is the *W*-line intensity compared to the intensities of the dielectronic satellites the higher is the temperature. The fit of the experimental spectrum presented in the inset of the Fig.4 was made using the generalized population kinetics and the spectral modeling code FLYCHK [36]. A steady state approximation that assumes constant in time and space plasma parameters gives a bulk electron temperature $T_e = 1250$eV. We found in the spectrum of *Li*- and *He*-like *Ti*-ions no influence of hot electrons [37]. This can be explained by the fact that the most fractions of hot electrons are generated at the maximum of the laser pulse intensity, while the plasma radiation occurs with 100fs delay (see simulation results in Fig. 8a).

The plasma electron density was determined based on the analysis of the relative intensities of the resonance ($W: 1s^2\ ^1S_0 - 1s2p\ ^1P_1$) and the intercombination ($Y: 1s^2\ ^1S_0 - 1s2p\ ^3P_1$) transitions in He-like Ti ions. It has to be mentioned that Ti ($Z_n = 22$) is the heaviest metal in the periodic table, for which the group of temperature dependent dielectronic satellites can be resolved from the intercombination line $Y$ used for density diagnostics [30, 38]. For Fe, Cu, Zn, etc. these lines are energetically mixed. Additionally, diagnostics that use the density-dependent ratio $\alpha$ of the resonance and intercombination lines can be applied only in a severely restricted density region, which is defined by the nuclear charge of the diagnostic element [38]. Figure 6 shows this dependence, e. g., for Si-, Ti- and Fe-plasmas calculated with the FLYCHK-code.

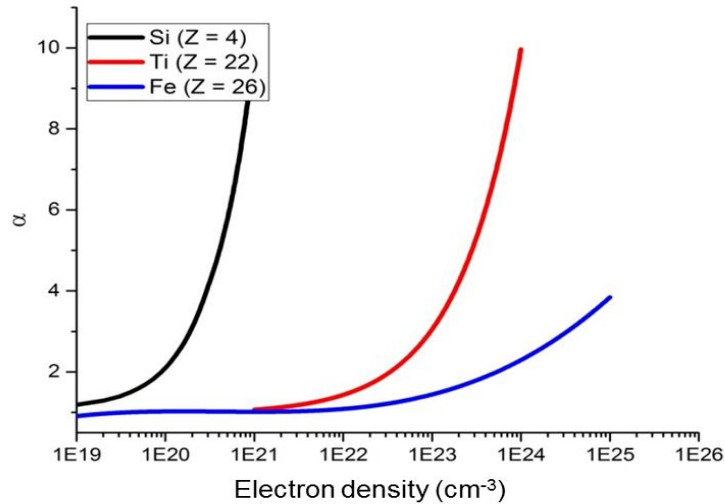

FIG. 6. Relative intensities $\alpha$ of the resonance $W$ ($1s^2\ ^1S_0 - 1s2p\ ^1P_1$) and intercombination $Y$ ($1s^2\ ^1S_0 - 1s2p\ ^3P_1$) transitions in *He*-like ions as a function of the electron density calculated using the FLYCHK code for Si-, Ti- and Fe-plasmas at temperatures $T_e = Ry \times Z_n^2/5$, $Ry=13.6$ eV.

From this figure it is obvious that Si would be suitable for density measurements between $10^{20}$cm$^{-3}$ and $10^{21}$cm$^{-3}$ and Fe above $10^{23}$cm$^{-3}$ (apart from the problem arising from the diagnostic lines mixing mentioned above). Thus, Titanium ($Z_n = 22$) is the best suited element for diagnostics of keV-hot plasmas at near-solid electron density.

Measured high intensity of the resonance line $W$ compared to the intercombination line $Y$ ($\alpha = 3-4$) indicates a high plasma electron density. Note that for the critical electron density of $n_e = 7 \times 10^{21}$cm$^{-3}$ corresponding to the 400 nm laser wavelength $\alpha = 1.25$. According to the fit made for the optically thin case and presented in Fig. 5, we obtain $n_e = 1.7 \times 10^{23}$ cm$^{-3}$ or 15% of the Ti ion solid density ($n_{solid} = 5.6 \times 10^{22}$cm$^{-3}$) with a mean ion charge $Z_i = 20$.

If the mean photon path $L_{ph}$ in the plasma is shorter than the plasma dimensions $R$, the plasma opacity has to be taken into account by introducing of the optical thickness $\tau = R/L_{ph}$ [36]. Absorption of a self-radiation by a 100-200 nm thick layer of keV Ti-plasma leads to the suppression of the resonance line intensity by a factor of $(1+\tau) \sim 1.4-2$. At the same time, the plasma remains optically thin for the intercombination line $Y$ because of the low oscillator strength of the spin-forbidden transition. For dielectronic satellites, which are a result of radiative transitions to low populated excited states $1s^2 2l$, plasma is also optically thin, since $\tau$ is proportional to the population of the ionic states responsible for the resonant absorption [36, 39]. All this result into slightly higher bulk plasma temperature and electron density as evaluated for the optical thin case.

However, transient spectral analysis has to be done, since the interaction of the 45 fs long laser pulse with the target, when the electron temperature is rising rapidly, is shorter than collisional ionization and collisional excitation times estimated for Ti $K$-shell transitions at the evaluated in the steady-state approximation plasma parameters. The dependence of the ion charge state on time and the correct description of forbidden transitions with "long" relaxation times such as the $Y$-line are of great importance for obtaining of reliable data on plasma electron temperature and density in transient case [39]. The temporal behavior of the bulk electron temperature was taken from the PIC-simulations and scaled in the amplitude and the FWHM to obtain the best coincidence between synthetic and measured spectra. For transient spectra simulations we varied the amplitude of the temperature "pulse" and its duration from 80 fs (time during which the laser pulse exceeded the intensity level of $10^{18}$ Wcm$^{-2}$) up to 120 fs. We obtained the best fit for 120 fs FWHM with the maximum of the bulk electron temperature $T_e = 1.8$ keV. The target density was kept constant at the level of $10^{22}$ at·cm$^{-3}$ over the whole time history, whereas the electron density was calculated according to the temporal evolution of the Ti charge caused by the collisional ionization process and reached $2.2 \times 10^{23}$ cm$^{-3}$ (see Fig. 6).

Figure 7a shows the time evolution of the electron temperature and density, which allowed for the best fit of the experimentally measured spectrum. One can see that the maximum of the $He$-like ion fraction (Fig. 7b) appears 80-100 fs after the temperature peak that corresponds to the maximum of the laser intensity.

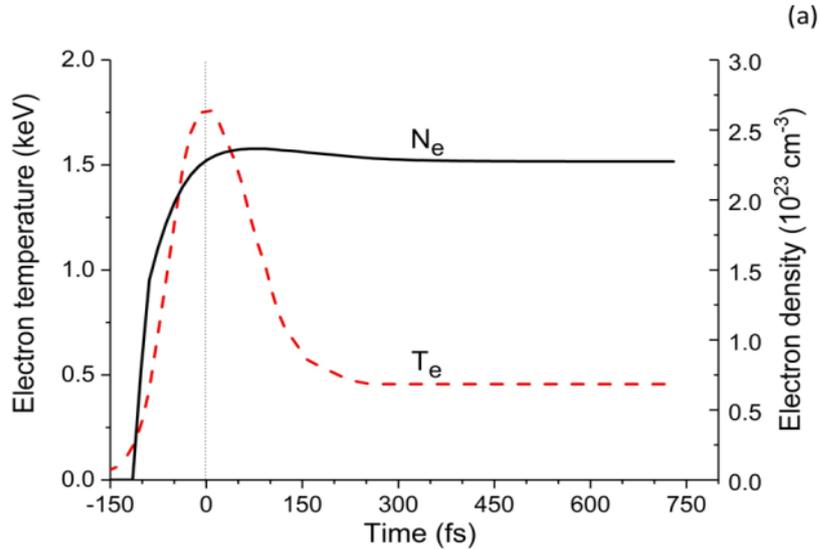

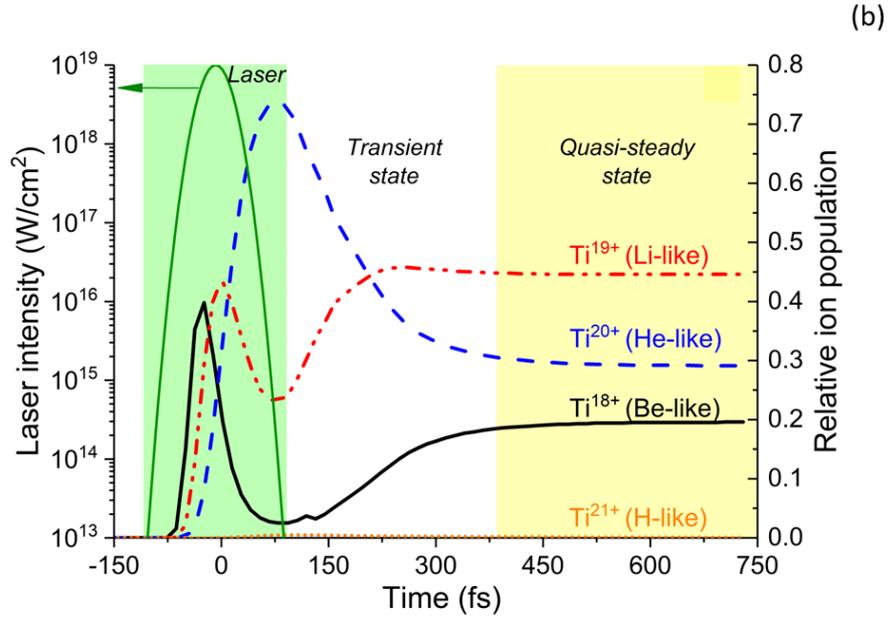

FIG. 7. Time history of the bulk electron temperature and density (a) providing the best fit of the measured spectrum and the corresponding evolution of the Ti-ion charge states (b).

Figure 8a shows the temporal evolution of the K-shell transitions in *Be*-, *Li*- and *He*-like Ti simulated with the transient version of the FLYCHK code. Here, *Be-sat* and *Li-sat* designates *K*-shell transitions $1s2l2l'2l''-1s^22l2l''$ ($l, l', l'' = s, p$) in double excited *Be*-like ($Ti^{18+}$) ions and $1s2l2l'-1s^22l$ ($l, l' = s, p$) in *Li*-like ions ($Ti^{19+}$) correspondently. The temporal behavior of the characteristic line intensities depends not only on the evolution of the plasma parameters, but also on characteristic time scales of collisional (ionization, bound electron excitation, etc.) and radiative processes, leading to the spectra formation.

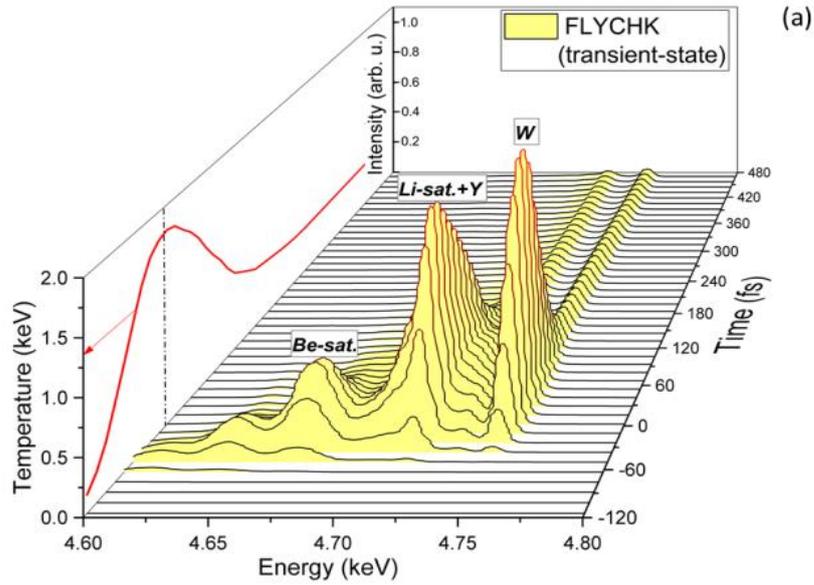

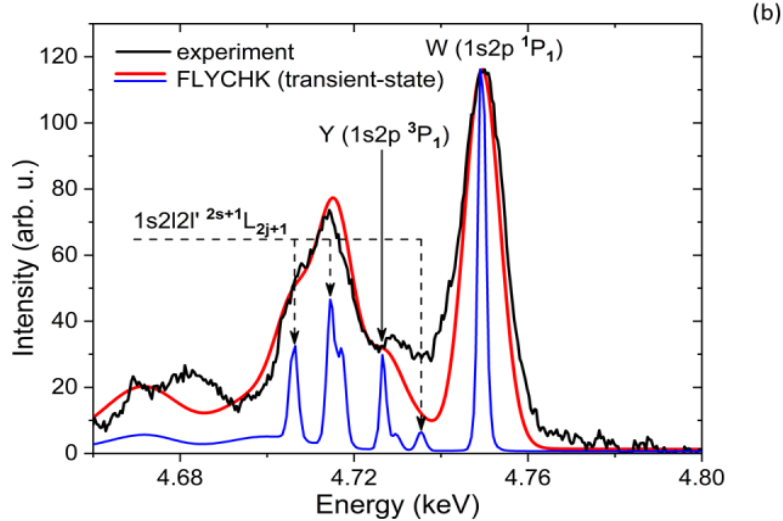

FIG. 8. Temporal evolution of the Ti characteristic radiation simulated with time dependable plasma parameters shown in Fig. 7a (a). Best fit of the experimental time integrated spectrum was obtained for a temperature "pulse" of 120 fs duration and 1.8 keV amplitude (b).

If the plasma density stays high during the time required to reach a steady state ion charge distribution and if this time is not shorter than the radiative relaxation time of metastable levels [39], then time integrated spectra can be treated in the steady-state approximation even for short scale interaction processes. If these conditions are not fulfilled, the steady-state treatment of time integrated spectra can lead to a strong misinterpretation of the plasma parameters [37, 39]. Fig. 8b shows the resulted time integrated synthetic spectrum (red solid line) together with the spectrum measured experimentally (black solid line). Blue lines demonstrate the spectral structure with narrow line widths in order to visualize the contribution of different line groups into the total spectrum. The result of the transient spectra fitting with the pick of the electron temperature at 1.8 keV (Fig. 7a) looks very similar to those made in the steady-state approximation for $T_e$=1.2 keV (see inset in Fig. 5). This can be explained by the high electron density and correspondingly high rates of collisions governing the ion charge development and population of the ion excited states The difference in the bulk electron temperatures is mainly caused by the fact that in the steady-state approximation the evaluated electron temperature presents the value that is averaged in time and space, while in the case of the transient procedure it is time resolved.

## V. EXPERIMENTAL MEASUREMENTS AND SIMULATIONS OF THE HOT PLASMA LAYER DEPTH.

The thickness of the hot plasma layer was investigated using 25 µm thick Ti-foils covered from the laser side by Fe-layers of various thicknesses from 0.05 up to 1 µm. The radiation of the hot Ti-plasma was registered below 50, 100 and 135nm thick Fe-layer. Starting from 200nm we could observe only $K_\alpha$ and $K_\beta$ - transitions in Ti (Fig. 9), a signature for the bulk electron temperature below 50-100 eV. Note that this type of radiation occurs when tens of keV hot electrons propagate into the "cold" target layers during the action of the laser pulse. Target heating in this case occurs both due to the electron collisional and resistive energy losses [1] and has a volumetric character. For estimations one can take the stopping path of electrons with energy 20-40 keV in cold Fe/Ti that results into 2-8 µm.

Characteristic radiation of Fe-plasma (see inset in Fig. 9) showed the presence of highly ionized $Fe^{24+}$ ions, the evidence for hot plasma states of the near solid density. Spectra-fitting in the steady-state approximation gives the electron temperature 1.5-1.6 keV. For the transient case, where the electron temperature is changing in time similar to the plot in Fig.7a, we receive a pick electron temperature ~2 keV. The difference in the temperatures evaluated from Ti and Fe $K$-shell spectra (assuming the same electron temperature of Fe and Ti-foils after action of the laser pulse) can be explained by the fact, that the minimum of the bulk electron temperature required for the $K$-shell spectra production depends on the nuclear charge of the diagnostic element ($T_e$ (eV)$\gtrsim Z_n^2 \times Ry/5$). Since the electron temperature is rapidly rising during the heating process and drops during plasma cooling, spectra of high $Z_n$ elements integrate time moments with higher temperatures [37].

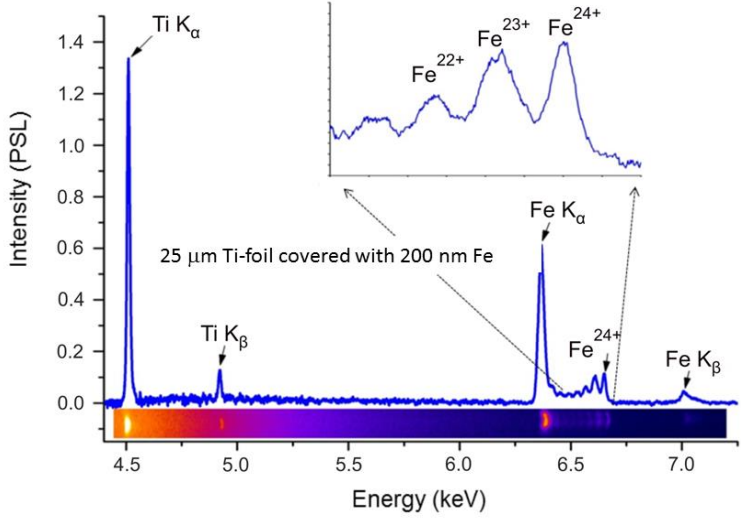

FIG. 9. K-shell radiation of the Ti-foil covered with a 200 nm thin Fe-layer. The spectrum shows "hot" K-shell transitions in *Be* -, *Li* - and *He*-like Fe-ions ($Fe^{+22}$ - $Fe^{+24}$, see inset) and "cold" $K_\alpha$ and $K_\beta$ radiation of Ti.

With increasing Fe-layer thickness one could observe two phenomena: the structure and relative intensities of the titanium *W*-line and dielectronic satellites, responsible for the electron temperature diagnostics, showed rather constant bulk electron temperature at the keV level, and the absolute intensities of Ti-lines dropped in a linear dependence with the hot Ti-layer thickness below the iron plasma if one supposes the total keV-hot Fe-Ti plasma thickness of 150nm.

Discussing the heating mechanisms that lead to generation of the keV hot dense plasma layer of 150 nm thickness we have consider first the contribution of supra-thermal electrons. Due to μms-long stopping path of these electrons they would produce rather volumetric energy deposition what is in contradiction with our measurements. Another mechanism that we propose to explain the experimental results is the target heating by thermal wave propagating from the vacuum-target interface and caused by the keV-hot bulk electrons presenting the major fraction and. These electrons are generated in the target skin layer with a near solid step-like electron density distribution due to collisional mechanism of the laser pulse absorption.

In order to prove the concept of a collisional heating of bulk electrons in the skin layer, the pilot 1D hydrodynamic (HD) simulations of the metal target dynamics from room temperatures to the conditions of weakly coupled plasma were performed. In simulations we used a wide-range model [7] with a two-temperature equation of state, a wide-range description of all transport and optical properties and took into account self-consistently the laser energy absorption in a target (Maxwell equations with a wide-range permittivity), ionization, electron and radiative heating, and plasma expansion.

Having in mind that in the skin layer an oscillating energy of electrons in the laser field is much higher than the keV-range temperature of bulk electrons, the approximation for the effective electron-ion collisional frequency $v_{eff}$ was used [41, 42]. This approximation provides a fitting expression in order to interpolate between two limits of weak and strong oscillating electric field in hot plasmas and reproduce results of analytical theories [43, 44]:

$$v_{eff} = v_{ei}(1 + v_E^2/6v_{T_e}^2)^{-3/2}$$

Here $v_{ei}$ is the electron-ion transport collision frequency that is invers proportional to the cube of the electron thermal velocity $v_{Te}$, and $v_E = eE/m_e\omega$ is the quiver electron velocity in a laser field of amplitude *E* and the frequency *ω*, *e* and $m_e$ are the charge and the mass of electron. This expression describes a strong reduction of the effective collision frequency with increasing of the electron oscillatory velocity $v_E$ above thermal electron velocity. It should be noted that a very high laser pulse contrast ensures a near solid step-like electron density in the skin layer and thus a strong reduction of the laser field in the absorption region that provides a nonrelativistic electron motion supposed in the above formula.

The HD simulations were performed for laser parameters described in Sec. II and the Fe-target. It was assumed that the target initially is the solid state matter at room temperature. Due to high laser contrast, the energetic of the pre-pulse stays below the plasma generation threshold in metals until the rising edge of the laser pulse reaches the target at $t = -50$ fs (note that in simulations, $t = 0$ fs corresponds to the pick of the laser intensity).

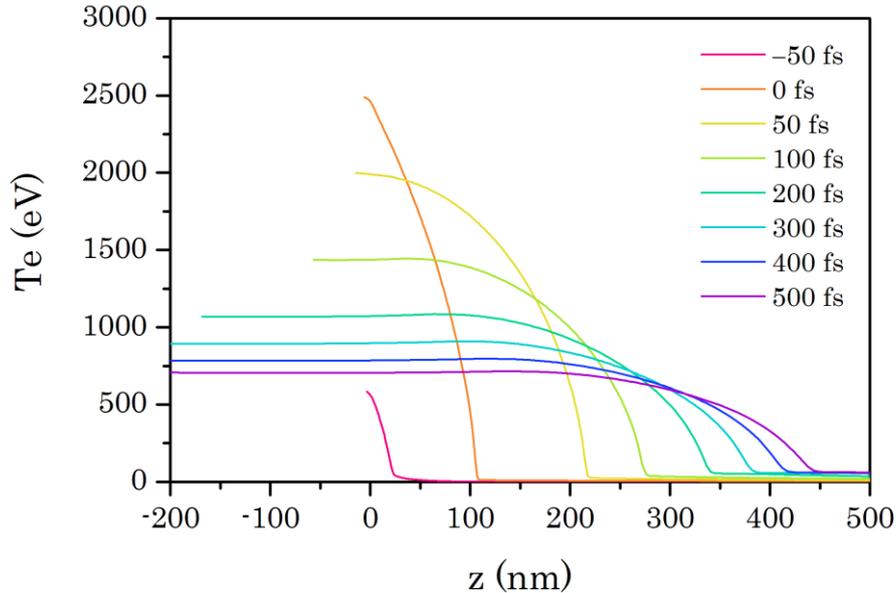

FIG. 10. Hydrodynamic simulations of the bulk electron temperature distribution in time and target depth.

In Fig.10, the spatial distribution of the bulk electron temperature is presented at different times up to 500 fs after the laser pulse maximum. The laser energy is deposited first into the 10-20 nm thin skin layer at $t = -50$ fs. To the maximum of the laser pulse intensity ($t = 0$ fs), the bulk electron temperature rises up to 2.5 keV and is transported into the first 100 nm of the target by the electron heat conductivity. By the end of the laser pulse ($t \sim 50$ fs) the target layer with temperature above 1.5 keV (this temperature is required for collisional excitation of $Fe^{+24}$) reaches 150 nm thickness. At later times ($t > 200$ fs), the deposited energy is redistributed over the thicker target layer and simultaneously drops below 1 keV.

Using evolution of the electron temperature obtained in HD-simulations, we can reconstruct a spatially and temporally resolved picture of the processes contributed into x-ray radiation. Part of the measured spectrum (Fig.9), which incorporates $K$-shell transitions in low ($K_\alpha$) and intermediate charged Fe-ions ($F$- up to $Be$-like) that exist in plasma at the bulk electron temperature below 1 keV, can be excited only by supra-thermal electrons. These lines contribute into the total spectra only during the laser pulse (-50fs ÷ +50fs), as long as the supra-thermal electrons are produced (see Fig.4). At $t = +50$ fs, the region with the electron temperature at the 1 keV level is moved up to 200 nm into the target depth (Fig. 10), so that electrons with energies > 20 keV and corresponding stopping range above 1 μm can produce $K$-shell ionization in "warm" and "cold" target regions.

The "hot" part of the measured spectra (inset in Fig. 9) that presents $Li$- and $He$-like Fe-ions ($Fe^{23+}$- $Fe^{24+}$) corresponds to the target regions with Te ≳ 1.5 keV, where according to HD-simulations, the electron temperature rises up to 2.5 keV at the maximum of the laser pulse and then drops down to the 1.5 keV level after 100fs at the depth of 150 nm (Fig. 10). Time resolved fitting of the hot Fe-plasma radiation gives the pick-temperature of 2 keV that is close to results of HD-simulations.

At the end of this section we would like to conclude that the bulk electron temperature profiles presented in Fig. 10 are in a good agreement with spectroscopic measurements of the heated target region determined with the help of Ti-foils covered by Fe-layers (see Section IV).

## VI. SUMMARY

In this work, we present the experimental evidence of the plasma states with the bulk electron temperatures above 1 keV and near-solid densities and analyze the mechanism of their generation.

The ultra-high energy density plasma states were produced due to interaction of the high contrast relativistic laser pulses of the doubled frequency with planar metallic foils. Bulk plasma parameters were investigated by measuring the characteristic radiation emitted from the target front side. The bulk electron temperature evaluated from the characteristic radiation of highly charged Ti-ions reached 1.8 keV and the plasma density 17% of the solid one assuming the homogeneous plasma. Electron temperature estimated from the characteristic transitions in $Fe^{23+}$- $Fe^{24+}$ showed up 1.5-1.6 keV for a steady-state approximation and ~2 keV for resoled in time electron temperature behavior. Note that plasma created during the short pulse laser-matter interaction is primarily transient, but extremely high electron density and correspondingly high rates of collisions governing the ion charge development and population of the ion excited states allow for application of a near steady-state description. Although the steady-state approach shows up a bit lower bulk electron temperatures compared to the more accurate transient approach, it doesn't required information on the temporal evolution of plasma parameters and can be used for prompt estimations of the plasma electron temperature providing its lower limit.

Using Ti-foils covered with nm-thin Fe-layers we have demonstrated that the thickness of the created keV hot dense plasma doesn't exceed 150nm. This value is in a good agreement with results of the pilot HD-simulations performed based on the wide-range two-temperature EOS, wide-range description of all transport and optical properties. According to these simulations, the generation of keV-hot bulk electrons is caused by the collisional mechanism of the laser pulse absorption in plasmas with a near solid step-like electron density profile. The laser energy firstly deposited into the nm-thin skin-layer is then transported into the target depth by the electron heat conductivity. This scenario is opposite to the volumetric character of the energy deposition produced by supra-thermal electrons that are responsible mostly for the heating of larger target volumes up to moderate temperatures below keV.

## ACKNOWLEDGMENTS

We thank B. Beleites, F. Ronneberger and JETI-team for running the laser system.

## REFERENCES

[1] A. Schoenlein, G. Boutoux, S. Pikuz, L. Antonelli, D. Batani, A. Debayle, A. Franz, L. Giuffrida, J. J. Honrubia, J. Jacoby, D. Khaghani, P. Neumayer, O. N. Rosmej, T. Sakaki, J. J. Santos and A. Sauteray, Europ. Phys. Lett. **114,** 45002 (2016).

[2] P. Neumayer, B. Aurand, M. Basko, B. Ecker, P. Gibbon, D. C. Hochhaus, A. Karmakar, E. Kazakov, T. Kuehl, C. Labaune, O. Rosmej, An.Tauschwitz, B. Zielbauer, and D. Zimmer, Physics of Plasmas **17,** 103103 (2010).

[3] T. Schlegel, S. Bastiani, L. Gre´millet, J.-P. Geindre, P. Audebert, and J-C. Gauthier, Phys. Rev. E **60**, N2 (1999).

[4] T. Liseykina, P. Mulser, and M. Murakami, Phys. Plasmas **22,** 033302 (2015).

[5] S. Atzeni, J Meyer-ter-Vehn, "Beam Plasma interaction, Hydrodynamics, Hot Dense Matter", International Series of Monographs on Physics 125, Oxford Science Publication (2004).

[6] O. Culfa, G. J. Tallents, E. Wagenaars, C. P. Ridgers, R. J. Dance, A. K. Rossall, R. J. Gray, P. McKenna, C. D. R. Brown, S. F. James, D. J. Hoarty, N. Booth, A. P. L. Robinson, K. L. Lancaster, S. A. Pikuz, A. Ya. Faenov, T. Kampfer, K. S. Schulze, I. Uschmann, and N. C. Woolsey, Phys. Plasmas **21**, 043106 (2014).

[7] M. E. Povarnitsin, N. E. Andreev, E. M. Apfelbaum, T. E. Itina, K.V. Khischenko, O. F. Kostenko, P. R. Levashov, M. E. Veysman, Appl. Surf. Sci. **258,** 9480–9483 (2012).

[8] M. E. Povarnitsin, N. E. Andreev, P. R. Levashov, K.V. Khischenko, D. A. Kim, V. G. Novikov, and O. N. Rosmej, Laser Part. Beams **31**, 663–671 (2013).

[9] L.P. Pugachev, N.E. Andreev, P.R. Levashov, O.N. Rosmej, Nucl. Instr. and Meth. A **829**, 88–93 (2016).

[10] N.E. Andreev, L. P. Pugachev, M. E. Povarnitsyn, and P. R. Levashov, Laser and Particle Beams **34**, 115–122 (2016).

[11] K.A. Ivanov, I.N. Tsymbalov, S.A. Shulyapov, D.A. Krestovskikh, A.V. Brantov, V.Yu. Bychenkov, R.V. Volkov, A.B. Savel'ev. Phys. Plasmas **24**, 063109 (2017).

[12] A.J. Kemp, F. Fiuza, A. Debayle, T. Johzaki, W.B. Mori, P.K. Patel, Y. Sentoku and L.O. Silva, Nucl. Fusion **54** 054002 (2014).

[13] P. Raczka, J.-L. Dubois, S. Hulin, V. Tikhonchuk, M. Rosinski, A. Zaras-Szydlowska, and J. Badziak, Laser Part. Beams , published online 2 November 2017; https://doi.org/10.1017/S026303461700074X

[14] V. Tikhonchuk, Phys. Plasmas **9**, 1416 (2002).

[15] J.J. Honrubia, C. Alfonsin, L. Alonso, B. Perez, J.A. Cerrada, Laser and Particle Beams **24**, 217–222 (2006).

[16] J. J. Honrubia,, M. Kaluza, J. Schreiber, G. D. Tsakiris, and J. Meyer-ter-Vehn, Phys. Plasmas **12**, 052708 (2005).

[17] V. Bagnoud and F. Wagner, High Power Laser Science and Engineering **4,** e39 (2016).

[18] M. A. Purvis, V. N. Shlyaptsev, R. Hollinger, C. Bargsten, A. Pukhov, A. Prieto, Y. Wang, B. M. Luther, L. Yin, S. Wang, and J. J. Rocca, Nature Photonics **7**, 796 2013 (2013).

[19] D. Khaghani, *Investigation of Laser Interaction with Nano-/Micro-Structures for Ion Acceleration, Intense X-Ray Production & High Energy Density Generation*, PhD 2016, Frankfurt University, https://www.gsi.de/en/work/research/appamml/plasma_physicsphelix/publications.htm

[20] D. Khaghani, M. Lobet B. Borm, L. Burr, F. Gaertner, L. Gremillet, L. Movsesyan, O. Rosmej, M. E. Toimil-Molares, F. Wagner, and P. Neumayer, Enhancing laser-driven proton acceleration by using tailored micro-pillar arrays at high drive energy, Scientific Reports **7**, 11366 (2017).

[21] C. Bargsten, R. Hollinger, M. G. Capeluto, V. Kaymak, A. Pukhov, S. Wang, A. Rockwood, Y. Wang, D. Keiss, R. Tommasini, R. London, J. Park, M. Busquet, M. Klapisch, V. N. Shlyaptsev, J. J. Rocca, Sci. Adv**. 3**, 1601558 (2017).

[22] Z. Samsonova, S. Hoefer, A. Hoffmann, B. Landgraf, M. Zürch, I. Uschmann, D. Khaghani, O. Rosmej, P. Neumayer, R. Roeder, L. Trefflich, C. Ronning, E. Foerster, C. Spielmann, and D. Kartashov, AIP Conf. Proceed. **1811**, 180001 (2017).

[23] A. P. L. Robinson, M. Zepf, S. Kar, R. G. Evans and C. Bellei, New Journal Phys.**10**, 013021 (2008).

[24] M. Kaluza, J. Schreiber, M. I. K. Santala, G. D. Tsakiris, K. Eidmann, J. Meyer-ter-Vehn, and K. J. Witte, Phys. Rev. Lett. **93**, 045003 (2004).

[25] B. Aurand, S. Kuschel, O. Jaeckel, C. Roedel, H. Y. Zhao, S. Herzer, A. E. Paz, J. Bierbach, J. Polz, B. Elkin, New Journal Phys. **15,** 033031 (2013).

[26] A. Saevert, S. P. D. Mangles, M. Schnell, E. Siminos, J. M. Cole, M. Leier, M. Reuter, M. B. Schwab, M. Möller, K. Poder, O. Jaeckel, G. G. Paulus, C. Spielmann, S. Skupin, Z. Najmudin, and M. C. Kaluza, Phys. Rev. Lett. **115**, 055002 (2015).

[27] C. Hahn, G. Weber, R. Maertin, S. Hoefer, T. Kaempfer, and Th. Stoehlker, Rev. Sci. Instrum. **87,** 043106 (2016).

[28] T. Bonnet, M. Comet, D. Denis-Petit, F. Gobet, F. Hannachi, M. Tarisien, M. Versteegen, and M. M. Aleonard, Rev. Sci. Inst. **84,** 103510 (2013).

[29] S. C. Wilks, W. L. Kruer, M. Tabak, and A. B. Langdon, Phys. Rev. Lett. **69**, 1383 (1992).

[30] Journal of Soviet Laser Research, **6**, N2 (1985) Editor N. G. Basov.

[31] M. V. Ammosov, N. B. Delone, and V. P. Krainov. Soviet Physics - JETP, **64**(6), 1191–1194 (1986).

[32] U. Zastrau, P. Audebert, V. Bernshtam, E. Brambrink, T. Kaempfer, E. Kroupp, R. Loetzsch, Y. Maron, Yu. Ralchenko, H. Reinholz, G. Roepke, A. Sengebusch, E. Stambulchik, I. Uschmann, L. Weingarten, and E. Foerster, Phys. Rev. E **81,** 026406 (2010).

[33] Ch. Reich, I. Uschmann, F. Ewald, S. Duesterer, A. Luebcke, H. Schwoerer, R. Sauerbrey, and E. Foerster Phys. Rev. E **68,** 056408 (2003).

[34] A. H. Gabriel, Monthly Notices of the Royal Astronomical Society **160,** 99–119 (1972).

[35] M. Schollmeier, G. R. Prieto, F. B. Rosmej, G. Schaumann, A. Blazevic, O. Rosmej, M. Roth, Laser Part. Beams **24**, 335-405 (2006).

[36] H.-K Chung., M. Chen, W. L. Morgan, Y. Ralchenko, R. W. Lee, High Energy Density Phys. **1**, N1, 3-12 (2005)

[37] F. B. Rosmej, O. N. Rosmej, 21st Conference on Control fusion and Plasma Physics, Monpellier, France, **3,** 1292-1295 (1994).

[38] A. V. Vinogradov, I. Yu Skobelev, E. A. Yukov. Sov. Jour. Quant. Electron. **5,** 630 (1975).

[39] O. N. Rosmej and F. B. Rosmej, Nucl. Instr. and Meth. B **98,** 37- 40 (1995).

[40] A. Pukhov, J. Plasma Phys. **61**, 425–433 (1999).

[41] R. J. Faehl and N. F. Roderick, Phys. Fluids **21**, 793 (1978).

[42] L. Schlessinger and J. Wright, Phys. Rev. A **20,** 1934 (1979).

[43] V. P. Silin, Soviet Phys. JETP **20**, 1510 (1965) [J. Exptl. Teoret. Phys. (U.S.S.R.) **47,** 2254 (1964)].

[44] L. A. Brantov, W. Rozmus, R. Sydora, C. E. Capjack, V. Yu. Bychenkov, V. T. Tikhonchuk, Phys. Plasmas **10**, No. 8, 3385 (2003)